\documentclass[%
11pt,
reprint,
onecolumn,
tightenlines,
superscriptaddress,
preprintnumbers,
nofootinbib,
amsmath,amssymb,amsthm,
physrev,
eqsecnum,tikz,
]{revtex4-2}

\usepackage{isomath}
\usepackage{amsmath,amsthm}
\usepackage{amsbsy}
\usepackage{amssymb}
\usepackage{amscd}
\usepackage{amsfonts}
\usepackage{stmaryrd}
\usepackage{siunitx}
\usepackage{euscript}
\usepackage[utf8]{inputenc}
\usepackage[T1]{fontenc}
\usepackage{newtxtext} 
\everymath{\displaystyle}
\usepackage{exscale}
\usepackage{microtype}
\usepackage{hyperref}
\usepackage{booktabs}
\usepackage{algorithm}
\usepackage{algpseudocode}

\usepackage{graphicx}
\usepackage{boxedminipage}
\usepackage{calc}
\usepackage[usenames,dvipsnames]{xcolor}
\graphicspath{ {media/} }
\usepackage[caption=false,justification=centerlast]{subfig}

\usepackage{setspace}
\usepackage{enumitem}
\setitemize{noitemsep,topsep=0pt,parsep=0pt,partopsep=0pt}
\setenumerate{noitemsep,topsep=0pt,parsep=0pt,partopsep=0pt}
\setdescription{noitemsep,topsep=0pt,parsep=0pt,partopsep=0pt}

\usepackage{orcidlink}
\usepackage{siunitx}
\usepackage[small]{titlesec}

\titlespacing*{\section}{0pt}{12pt plus 4pt minus 2pt}{2pt plus 2pt minus 2pt}
\titlespacing*{\subsection}{0pt}{12pt plus 4pt minus 2pt}{2pt plus 2pt minus 2pt}
\titlespacing*\subsubsection{0pt}{12pt plus 4pt minus 2pt}{2pt plus 2pt minus 2pt}
\titlespacing*\paragraph{0pt}{12pt plus 4pt minus 2pt}{2pt plus 2pt minus 2pt}

\makeatletter

    \renewcommand*{\p@subsection}{}
    
    \renewcommand*{\p@subsubsection}{}
\makeatother

\usepackage{isomath}
\usepackage{amsmath}
\usepackage{amssymb}
\usepackage{amscd}
\usepackage{amsfonts}
\usepackage{upgreek}

\theoremstyle{definition}

\AtEndEnvironment{remark}{\null\hfill\qedsymbol}

\AtEndEnvironment{example}{\null\hfill\qedsymbol}

\newcommand{\bfsigma}{\mathbold {\sigma}}

\newcommand{\parderiv}[2]{\frac{\partial #1}{\partial #2}}
\newcommand{\dm}{\ \mathrm{d}}
\newcommand{\deriv}[2]{\frac{\dm #1}{\dm #2}}

\newcommand{\bfF}{{\mathbold F}}

\newcommand{\bfI}{{\mathbold I}}

\newcommand{\bfT}{{\mathbold T}}

\usepackage{lipsum}

\newcommand{\Fp}{\bfF_i\left(\rho_0\right)}
\newcommand{\iFp}{\bfF_i^{-1}\left(\rho_0\right)}

\begin{document}


\preprint{Journal of Applied Mechanics, Vol. 91, 014501, 2024. \url{https://doi.org/10.1115/1.4062967}}

\title{Deformation Decomposition versus Energy Decomposition for Chemo- and Poro- Mechanics}

\author{Janel Chua}
    \email{janelchua@gmail.com}
    \affiliation{Department of Civil and Environmental Engineering, Carnegie Mellon University}

\author{Mina Karimi}
    \affiliation{Department of Mechanical and Civil Engineering, California Institute of Technology}

\author{Patrick Kozlowski}
    \affiliation{Department of Civil and Environmental Engineering, Carnegie Mellon University}

\author{Mehrdad Massoudi}
    \affiliation{National Energy Technology Laboratory, 626 Cochran Mill Road Pittsburgh, PA 15236}%

\author{Santosh Narasimhachary}
    \affiliation{Siemens Corporation}

\author{Kai Kadau}
    \affiliation{Siemens Energy Inc}

\author{George Gazonas}
    \affiliation{DEVCOM Army Research Laboratory, Aberdeen Proving Ground, 21005 MD, USA}

\author{Kaushik Dayal}
    \affiliation{Department of Civil and Environmental Engineering, Carnegie Mellon University}
    \affiliation{Center for Nonlinear Analysis, Department of Mathematical Sciences, Carnegie Mellon University}
    \affiliation{Department of Mechanical Engineering, Carnegie Mellon University}

\date{\today}


\begin{abstract}
    We briefly compare the structure of two classes of popular models used to describe poro- and chemo- mechanics wherein a fluid phase is transported within a solid phase.    
    The multiplicative deformation decomposition has been successfully used to model permanent inelastic shape change in plasticity, solid-solid phase transformation, and thermal expansion, which has motivated its application to poro- and chemo- mechanics.
    However, the energetic decomposition provides a more transparent structure and advantages, such as to couple to phase-field fracture, for models of poro- and chemo- mechanics.
\end{abstract}

\maketitle


\section{Introduction}

There is significant current interest in modeling problems of fluid transport in porous media as well as fluid phase transport in solid materials, i.e., poro- and chemo- mechanics.
The motivations range from modeling hydrogels \cite{hong2008theory}, to transport in geological structures \cite{karimi2022energetic}, to hydrogen embrittlement of metals \cite{martinez2018phase}, among various other applications.
An approach that has been proposed in the literature to model poro- and chemo- mechanics is to decompose the deformation gradient into an elastic part -- that causes stress -- and an inelastic part -- that accounts for the shape change due to fluid transport; this is the ``multiplicative decomposition''.
The application of the multiplicative decomposition to poro- and chemo- mechanics is motivated by the success of this strategy in modeling thermoelasticity, plasticity, twinning, solid-solid phase transformations, and related phenomena that involve inelastic deformation, e.g. reviewed in \cite{lubarda2004constitutive,sadik2017origins}.

However, an important distinction between thermoelasticity, plasticity, twinning on the one hand, and poro- and chemo- mechanics on the other hand, is that the former class of phenomena do not involve the introduction of material into the bulk of the solid, whereas the latter class do.
The introduced material has energy and stress that is distinct from the energy and stress of the solid. 
This motivates an approach that is based on additively combining the energies of the solid and the fluid, e.g. \cite{hong2008theory,von2021morphogenesis,truster2017unified} and many others.

In this note, we briefly contrast the overall structure of these two approaches, and argue that the additive decomposition of the energy is the more appealing alternative.
We also highlight \cite{chenchiah2014energy}, which critically examined a model micromechanism for biological growth, and consequently argued against a multiplicative decomposition in that context.

\paragraph*{Definitions and Notation.}

We use $\bfF$ for the deformation gradient, $\bfT$ for the 1st Piola-Kirchoff (P-K) stress, and $\mu$ for the chemical potential.
For simplicity, we assume a single fluid phase that is defined by the densities in the deformed and the reference configurations $\rho$ and $\rho_0$; i.e., $\rho$ and $\rho_0$ are the mass of the fluid phase per unit deformed and reference volumes.
For simplicity, we follow the affine deformation assumption that the volume of the fluid phase in the deformed and reference configurations are related by $J = \det \bfF$, implying that $\rho = J^{-1} \rho_0$.

The energy density is written in terms of $\bfF$ and $\rho_0$, rather than $\bfF$ and $\rho$, because the former pair of arguments can be independently varied in a simple way that decouples deformation and transport.

\section{Multiplicative Deformation Decomposition into Elastic and Inelastic Parts}
\label{sec:strain}

The central idea in the multiplicative deformation decomposition is to write the deformation gradient $\bfF$ as the product of an elastic part $\bfF_e$, that causes stress, and an inelastic part $\Fp$, that is driven by the coupled field $\rho_0$.
That is, $\bfF = \bfF_e \bfF_i$, and the free energy density is typically of the general form given by:
\begin{equation}
    W(\bfF,\rho_0)
    =
    W_e \left( \bfF\iFp \right)
    +
    W_i(\rho_0)
\end{equation}
The elastic energy $W_e$ is minimized when $\bfF = \Fp$ up to rotations.

The resulting P-K stress has the form:
\begin{equation}
    \bfT = \parderiv{W}{\bfF} = \parderiv{W_e}{\bfF_e} \bfF_i^{-T}
\end{equation}
In general, $\bfF_i$ is invertible.
Consequently, $\bfT = {\bf 0} \iff \parderiv{W_e}{\bfF_e} = {\bf 0}$.

The (referential) chemical potential is the key quantity that governs the transport of the fluid phase.
It is defined as the energy-conjugate to $\rho_0$, e.g. \cite{coussy2004poromechanics}, and has the form:
\begin{equation}
\label{eqn:mu-multi-decomp}
    \mu 
    = \parderiv{W}{\rho_0} 
    = \parderiv{W_e}{\bfF_e} \deriv{\bfF_i^{-T}}{\rho_0} : \bfF + \deriv{W_i}{\rho_0}
\end{equation}
where $:$ represents a double contraction over 2-nd order tensors.

We note the key undesirable features of this class of models.
Consider a homogeneous body described by such a model with zero applied traction on the entire boundary and uniform $\rho_0$.
A solution to this boundary-value problem is $\bfT \equiv {\bf 0}$, implying that $\bfF=\bfF_i$ on the entire body, up to a rigid rotation.
Hence, even if the solid material is highly deformed due to fluid infiltration, e.g. due to hydrogen in a metallic lattice with stretched atomic bonds or due to fluid in a hydrogel with stretched polymer chains, the elastic energy $W_e$ is minimized.
While it is possible to augment the inelastic energy $W_i$ to depend on the deformation, this would not allow the interpretation of the decomposition of $\bfF$ as an elastic and inelastic part.

\section{Additive Energy Decomposition into Solid Strain Energy and Fluid Energy}
\label{sec:energy}

The central idea in the additive energy decomposition is to additively combine the energetic contributions of the solid and fluid phases to find the total free energy.
An example of such a form is:
\begin{equation}
\label{eq:W-add-decomp}
    W(\bfF,\rho_0) = \alpha W_s(\bfF) + (1-\alpha) J W_f\left(J^{-1}\rho_0\right)
\end{equation}
where $J=\det \bfF$.
The referential volume fraction of the solid phase is $\alpha$, and we assume a single fluid phase; for the case with multiple fluids with the possibility of evolving volume fractions, we refer to \cite{karimi2022energetic} and references therein.

The form of the fluid contribution $J W_f\left(J^{-1}\rho_0\right)$ is motivated by the requirement that the energy density $W_f$ of a simple fluid depends only on the density in the deformed state, i.e., $\rho = J^{-1} \rho_0$, when we consider the isothermal setting.
Further, the leading factor of $J$ accounts for the fact that $W_f$ is the energy per unit deformed volume, whereas the hyperelastic energy density $W$ is per unit reference volume.
An important assumption here is that of affine deformation; i.e., both the solid skeleton and the fluid volume deform under $\bfF$ affinely but this can be relaxed \cite{karimi2022energetic}.

The resulting P-K stress has the form:
\begin{equation}
\begin{split}
    \bfT
    & = \parderiv{W}{\bfF}
    = \alpha\parderiv{W_s}{\bfF} + (1-\alpha) \parderiv{J}{\bfF} \left(W_f\left(J^{-1}\rho_0\right) - J^{-1} \rho_0 \parderiv{W_f}{\rho}\right)
    = \alpha\parderiv{W_s}{\bfF} + (1-\alpha)J \bfF^{-T} \left(W_f\left(\rho\right) -  \rho \parderiv{W_f}{\rho}\right)
    \\
    & = \bfT_s + (1-\alpha)J \bfF^{-T} p
\end{split}
\end{equation}
where we have defined the solid stress $\bfT_s := \alpha\parderiv{W_s}{\bfF}$;
used the relation $\parderiv{J}{\bfF} = J \bfF^{-T}$;
and used the relation that the fluid pressure\footnote{
    The pressure $p$ is the derivative of the Helmholtz free energy with respect to volume, keeping temperature and mass fixed \cite{gurtin1982introduction}.
    In terms of the density $\rho$ -- which is inversely proportional to the volume when the mass is fixed -- and in terms of the Helmholtz free energy \emph{density}, we have:
    \begin{equation}
        p = \parderiv{\ }{\left(\frac{1}{\rho}\right)}\left(\frac{1}{\rho}W_f(\rho)\right) = W_f(\rho) - \rho \parderiv{W_f}{\rho}
    \end{equation}
} is given by $p=\left(W_f\left(\rho\right) -  \rho \parderiv{W_f}{\rho}\right)$.
We can then define the P-K fluid stress $\bfT_f := (1-\alpha) J \bfF^{-T}p$, corresponding to a Cauchy stress $\bfsigma_f = (1-\alpha) p \bfI$.

The chemical potential for this model has the form:
\begin{equation}
\label{eqn:mu-add-decomp}
    \mu = \parderiv{W}{\rho_0} = (1-\alpha)\parderiv{W_f}{\rho}
\end{equation}
which corresponds to the standard thermodynamic expression for fluids.

We consider again a homogeneous body with zero applied traction on the entire boundary and uniform $\rho_0$.
In this energetic decomposition model, a solution to this boundary-value problem is that the total stress $\bfT \equiv {\bf 0}$, implying that the fluid and solid stresses $\bfT_s$ and $\bfT_f$ balance each other but neither is necessarily zero.
Given a fluid pressure $p\neq 0$, there will generally be a fluid stress $\bfT_f \neq {\bf 0}$ which in turn requires a solid stress $\bfT_s \neq {\bf 0}$.
With this state of fluid and solid stress, $W_s$ will not reach its minimum, and the body will deform.
Hence, the deformation of the solid material due to fluid infiltration, e.g. the stretching of atomic bonds or polymer chains, will be reflected in the solid stress $\bfT_s$ and energy $W_s$.

\section{A Remark on Phase-field Fracture Modeling for Poro- and Chemo- Mechanics}

The phase-field approach provides a powerful method for modeling fracture, e.g. \cite{clayton2014geometrically}.
Briefly, a phase-field $\phi$ tracks the level of damage, with $\phi=1$ denoting the intact undamaged material and $\phi=0$ denoting the completely damaged or fractured material.
An energetic framework uses an energy density with contributions that include $\phi^2 W(\bfF) + G_c (1-\phi)^2$, where the first term accounts for the elastic energy and the second for the work to fracture, with $G_c$ being the Griffith parameter.
This structure of the energy density sets up a competition between elastic energy and the work to fracture: minimizing over $\phi$ drives $\phi \to 0$ when the elastic energy becomes larger  due to deformation than the work to fracture.
Given this reasoning, it is natural to develop poro- and chemo- mechanical models of phase-field fracture wherein only the energy of the solid phase $W_s$ contributes to the fracture energetic balance.
That is, in a simple model, we would replace \eqref{eq:W-add-decomp} by the expression $\phi^2 W_s(\bfF) + J W_f\left(J^{-1}\rho_0\right) + G_c (1-\phi)^2$ to model fracture which releases the stress in the solid but does not directly affect the fluid.


\paragraph*{Competing Interest Statement.}

The authors have no competing interests to declare.

\paragraph*{Acknowledgments.}

We acknowledge financial support from ARO (MURI W911NF-19-1-0245) and NSF (DMREF 2118945, DMS 2108784); NSF for XSEDE computing resources provided by Pittsburgh Supercomputing Center; and Noel Walkington and Tony Rollett for useful discussions.
Kaushik Dayal acknowledges an appointment to the National Energy Technology Laboratory sponsored by the U.S. Department of Energy.



\end{document}